\documentclass{article}
\usepackage{spconf,amsmath,graphicx,hyperref}
\usepackage{multirow}
\usepackage{booktabs}
\usepackage{tcolorbox}
\usepackage{makecell}
\usepackage{graphicx}
\usepackage[table]{xcolor}

\title{SEEM: Exploiting Black-Box Text Attacks to Manipulate Tool Selection}
\name{Author(s) Name(s)\thanks{Thanks to XYZ agency for funding.}}
\name{Liuji Chen$^{1,2,3,4,*}$, Hao Gao$^{5,*}$, Jinghao Zhang$^{1,2,3,4}$, Qiang Liu$^{1,2,3,4,{\dagger}}$, Shu Wu$^{1,2,3,4}$, Liang Wang$^{1,2,3,4}$
\thanks{$^{*}$Equal contribution.}
\thanks{$^{\dagger}$Corresponding author.}}
\address{$^1$New Laboratory of Pattern Recognition (NLPR) \\
$^2$ State Key Laboratory of Multimodal Artificial Intelligence Systems \\
$^3$ Institute of Automation, Chinese Academy of Sciences \\
$^4$ School of Artificial Intelligence, University of Chinese Academy of Sciences \\
$^5$ Beijing University of Posts and Telecommunications
}

\begin{document}
%
\maketitle
\begin{abstract}
Tool learning has emerged as a powerful auxiliary mechanism that extends the capabilities of large language models (LLMs), enabling them to address complex tasks that demand real-time relevance or high-precision operations. However, beneath this strength lie significant security risks. Prior studies have primarily concentrated on corrupting the outputs of invoked tools, while largely overlooking the vulnerability of the tool selection process itself. To bridge this gap, we introduce a black-box, text-based attack that substantially increases the likelihood of a target tool being selected. We propose SEEM, a two-level coarse-to-fine perturbation method that operates at both the word and character levels. Through comprehensive experiments, we show that merely perturbing the textual information of tools can markedly raise the probability of the target tool being prioritized and ranked higher among candidates. Our findings expose critical weaknesses in the tool selection mechanism and lay the groundwork for developing defenses to secure this essential process.
\end{abstract}
\section{Introduction}
\label{sec:intro}
In recent years, large language models (LLMs) such as ChatGPT, LLaMA, and Claude have demonstrated remarkable capabilities across diverse tasks. Yet, they remain limited in handling high-precision reasoning and real-time information integration. To address this, tool learning has been proposed, enabling LLMs to invoke external tools for enhanced accuracy and utility. This workflow typically involves four stages: task planning, tool selection, tool invocation, and response generation.

\begin{figure}
    \centering
    \includegraphics[width=\linewidth]{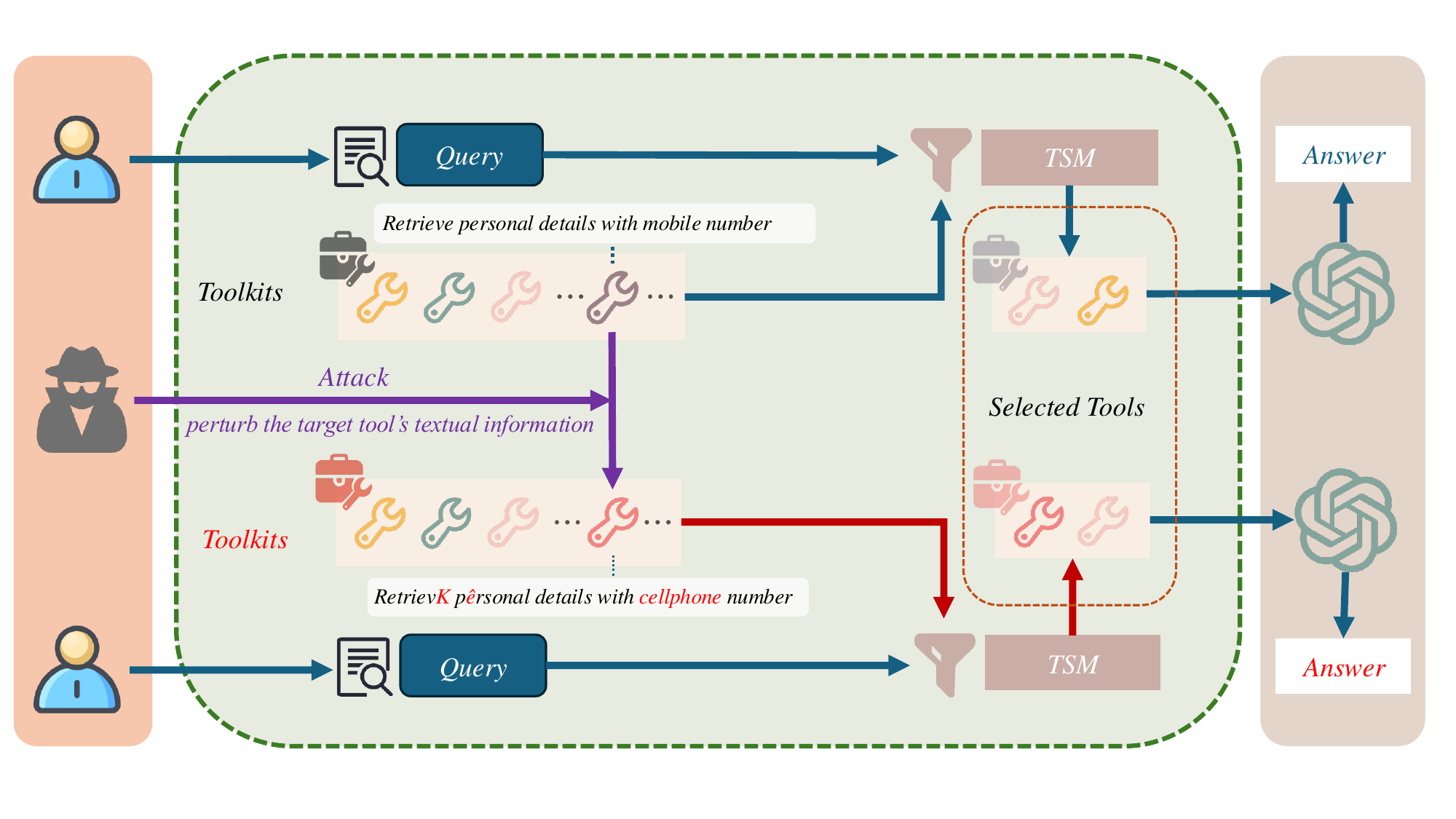}
    \caption{Overview of the black-box text attack on tool selection. The attacker perturbs the target tool’s textual information and identifies an adversarial text, which misleads the tool selection model \textit{TSM} into selecting the target tool, an outcome that would not have occurred otherwise. The modified parts of the target tool’s text are highlighted in red.}
    \label{fig:overview}
\end{figure}

While prior work has focused on the latter stages, emerging studies reveal that tool learning introduces new security risks. Existing attacks mainly target malicious tool invocation or harmful response generation, leaving the tool selection stage largely unexplored. However, manipulating this stage can serve two major attacker motivations: (i) maximizing commercial profits by biasing competition among monetized tools, and (ii) increasing the likelihood of malicious tools being invoked. In both cases, the attacker’s success hinges on ensuring their designated tool is selected as often as possible, as shown in Figure \ref{fig:overview}.

To bridge this gap, we propose the first black-box text attack on tool selection, termed SEEM. By applying coarse-to-fine perturbations at both word and character levels, SEEM subtly alters the textual metadata of target tools to mislead the selection model, thereby elevating the target tool’s ranking without impairing its functionality. This work highlights overlooked vulnerabilities in tool learning and provides systematic evaluation across both retriever- and LLM-based selection models.

\section{Related works}
\label{sec:related}
\subsection{Tool Selection}
Tool selection aims to identify the most suitable tools from a toolkit to solve user queries. Existing research falls into three paradigms: LLM-based, Generative, and Retriever-based selection. LLM-based methods exploit in-context learning for query understanding but are constrained by context window size, making them suitable only for small toolkits\cite{gao2023confucius, shen2023hugginggpt, lu2023chameleon}. Generative methods map tools to special tokens, enabling direct prediction during inference\cite{toolkengpt, wang2024toolgenunifiedtoolretrieval}; however, they are resource-intensive, require fine-tuning, and lack adaptability to dynamic tools, limiting real-world applicability. Therefore, this approach is not practical for real-world applications, and this paper will not discuss this paradigm. Retriever-based methods match queries to tools and rank candidates, with approaches ranging from BM25 + GPTIndex \cite{patil2023gorilla} to Sentence-BERT retrievers for efficient tool retrieval \cite{qin2023tool}.
\subsection{Text Attack}
Adversarial attack on discrete data such as text is more challenging than on continuous data, especially under black-box condition. Inspired by heuristic algorithms, TextBugger\cite{li2019textbugger} uses a genetic algorithm to generate adversarial examples by modifying input text in a way that misleads the model. In word-level perturbations, TextFooler\cite{jin2020textfooler} performs perturbations by focusing on semantically similar word replacements to create adversarial text.

\section{Threat Model}
\label{sec:threatmodel}
We consider a black-box adversary whose goal is to maximize the selection probability of a target tool. The attacker has no access to model parameters or internal states of the tool selection system, but can query it and observe outputs in the form of rankings or probabilities. The only capability assumed is the ability to modify the textual metadata (title, description, manual) of the tool they control.

The attack is subject to three constraints: (i) modifications must preserve semantic plausibility and readability to ensure the tool remains functional; (ii) the perturbation budget is limited; (iii) perturbations should remain stealthy, resembling natural variations such as synonyms or typos.

This setting reflects realistic scenarios: tool providers or malicious actors can easily update metadata to bias retrieval or LLM-based selection. Such attacks can be used either to maximize commercial profit (increasing invocation of benign tools) or to amplify the impact of malicious tool injection.

\section{Method}
\label{sec:method}
In this section, we introduce SEEM (\textbf{S}\textbf{e}l\textbf{e}ct \textbf{M}e! When you need a tool), a coarse-to-fine black-box adversarial attack framework for tool selection models. SEEM is designed to manipulate the textual information of a target tool such that its probability of being selected increases, without access to model parameters or gradients.

\subsection{Problem Formulation}
Let $\mathcal{T} = \{T_1,...,T_m\}$ be a set of tools, each with textual metadata $d_i$ (title, description, manual). Given a query $q$, the tool selection model (\textit{TSM}) computes a probability distribution:
\begin{align}
    \mathcal{P} = \textit{TSM}(q, \mathcal{D}), \quad \mathcal{D} = \{d_1, d_2, \dots, d_m\},
\end{align}
where $p_i$ denotes the selection probability of tool $T_i$. Our goal is to increase the probability of a target tool $T_t$ by perturbing its text $d_t$ into an adversarial version $d_t'$. The attack is black-box, assuming only query access to the tool selection model with outputs being either the full distribution $\mathcal{P}$ or top-ranked predictions (Section \ref{sec:threatmodel}).

Formally, SEEM solves:
\begin{align}
    \max_{d_t'} \; p_t(q; d_t', \mathcal{D}_{-t}),
\end{align}
where $\mathcal{D}_{-t}$ denotes the set after removing the target tool $T_t$.

\subsection{Stage I: Word-Level Attack}
First, coarse-grained perturbations, specifically at the word level, are applied to the textual information for the target tool. This entire perturbation process is conducted in a black-box setting, without guidance from any internal parameters or gradient information. Therefore, to efficiently find the target adversarial text, SEEM opts to start with words that have a more significant impact on the tool selection outcome. SEEM first identifies important words in the tool text. For each word $w_i$ we mask or delete it to obtain $d_t^{(-i)}$ and compute:
\begin{align}
\label{eq:imp}
    I(w_i) = p_t(q; d_t, \mathcal{D}_{-t}) - p_t(q; d_t^{(-i)}, \mathcal{D}_{-t})
\end{align}

The importance of a word is measured by the magnitude of $I(\cdot)$. Under a constrained attack budget, more important words are preferentially perturbed. If probability outputs are unavailable, we approximate importance using repeated winner-switch counts under ranking feedback.

For each individual word, SEEM introduces perturbations in several possible ways: deletion, swapping with an adjacent word, substitution with a synonymous term, or insertion of a semantically similar word. These operations, while searching for adversarial texts, preserve the original semantic content to the greatest extent possible, thereby ensuring a degree of stealthiness in the attack.

In this stage, we employ greedy search. At each step, SEEM queries the tool selection model with candidate modifications, selects the one maximizing $p_t$, and updates the text until the budget $\mathcal{B}_w$ is reached and get the coarse adversarial text $d_t^{(1)}$.

\subsection{Stage II: Character-Level Refinement}
After performing coarse-grained word-level perturbations, SEEM further
introduces fine-grained character-level perturbations to bias the tool
selection model.
Similar to the word-level stage, SEEM first estimates the importance of
each word using Eq.~\ref{eq:imp}, and then perturbs the characters within
each word in descending order of their importance.

At the character level, SEEM considers four perturbation operations:
\emph{insertion}, \emph{substitution}, \emph{deletion}, and
\emph{transposition}.
These operations are designed to preserve the readability of the text
while introducing subtle changes that influence the model’s decision.

To explore the perturbation space more effectively and avoid poor local
optima, SEEM employs a beam search strategy instead of a greedy update.
At each iteration, the algorithm maintains the top-\(k\) candidate
adversarial texts ranked by the target probability \(p_t\).
For each candidate, all feasible character-level perturbations are
generated and evaluated, and the beam is updated with the best-performing
candidates.
This process continues until the character-level perturbation budget
\(\mathcal{B}_c\) is exhausted or the search converges.

The final adversarial text, denoted as \(d_t^{*}\), combines
semantic-preserving word-level perturbations with subtle character-level
refinements, making it effective at manipulating tool selection while
remaining difficult to detect.
\begin{table}[]
\centering
\resizebox*{\linewidth}{!}{
\begin{tabular}{@{}ccccccc@{}}
\toprule
\multirow{2}{*}[-0.25em]{Model} 
           & \multicolumn{2}{c}{Hit@1}     & \multicolumn{2}{c}{Hit@3}     & 
           \multicolumn{2}{c}{Hit@5}       \\ \cmidrule(l){2-3} \cmidrule(l){4-5} \cmidrule(l){6-7} 
           & Origin  & Attack   & Origin  & Attack  & Origin  & Attack    \\ \midrule 
BM25
& 0.000  & 1.330  
& 0.195  & 3.535  
& 0.320  & 6.105      \\
Ada
& 0.000  & 2.686   
& 0.008  & 5.214   
& 0.231  & 7.900     \\
API-retriever  
& 0.052  & 1.814  
& 0.180  & 4.122  
& 0.836  & 5.628     \\
\bottomrule
\end{tabular}
}
\caption{Performance (in per thousand) of indiscriminate attack on different retrieval models at ToolBench I1.}
\label{tab:G1Res}
\end{table}

\begin{table}[]
\centering
\resizebox*{\linewidth}{!}{
\begin{tabular}{@{}ccccccc@{}}
\toprule
\multirow{2}{*}[-0.25em]{Model} 
           & \multicolumn{2}{c}{Hit@1}     & \multicolumn{2}{c}{Hit@3}     & 
           \multicolumn{2}{c}{Hit@5}       \\ \cmidrule(l){2-3} \cmidrule(l){4-5} \cmidrule(l){6-7} 
           & Origin  & Attack    & Origin  & Attack  & Origin  & Attack    \\ \midrule 
BM25
& 9.93  & 70.51  
& 34.11  & 89.01  
& 57.04  & 93.72      \\
Ada
& 12.89  & 59.11  
& 53.39  & 84.27  
& 82.29  & 96.38      \\
API-retriever  
& 12.44  & 45.53  
& 38.64  & 81.35  
& 66.33  & 93.00      \\
\bottomrule
\end{tabular}
}
\caption{Performance (in percentage) of conditional attack on different retrieval models at ToolBench I3.}
\label{tab:G3Res}
\end{table}

\section{Experiments}
\label{sec:experiments}
\begin{table}[]
\centering
\resizebox*{0.6\linewidth}{!}{
\begin{tabular}{@{}cccc@{}}
\toprule
\multirow{2}{*}[-0.25em]{Model} 
           & \multicolumn{3}{c}{\(\mathcal{P}_{\textit{use}}\)}          \\ \cmidrule(l){2-4} 
           & Origin  & Attack  &  Impro.  \\ \midrule 
Llama-3.2-3B
& 0.2571  & 0.7001  & 172\%    \\		
Qwen-2.5-7B
& 0.2594  & 0.4278  & 65\%    \\		
Vicuna-7B
& 0.0519  & 0.9929  & 1812\%    \\
Llama-8B
& 0.3491  & 0.9528  & 173\%    \\
Phi-4-14B
& 0.1971  & 0.4116  & 109\%    \\ 			
GPT-3.5  
& 0.3700  & 0.4600  & 24\%    \\
GPT-4o
& 0.2229  & 0.2988  & 34\%    \\ 		
\bottomrule
\end{tabular}
}
\caption{Performance of LLM-based selection attack on different LLMs at ScienceQA, where Impro. denotes relative improvement against origin results for target tool.}
\label{tab:llmRes}
\end{table}
\subsection{Experimental Setup}
To evaluate the effectiveness of the proposed method, two benchmark datasets are adopted to our experiments. Specifically, we employ ToolBench\cite{qin2023toolllm}, which integrates over 16k real-world APIs, including tools from more than 40 categories such as movies, sports, food, and more, and ScienceQA\cite{lu2022learn}, which consists of 21k multimodal multiple choice questions with diverse science topics and annotations of their answers with corresponding lectures and explanations. For the retriever-based experiments, we select BM25\cite{bm25}, OpenAI’s text-embedding-ada-002, and API-retriever\cite{qin2023toolllm} as the victim models. For the LLM-based experiments, we choose several models with different parameter scales, including Llama-3.2-3B, Qwen-2.5-7B\cite{qwen2_technical_report}, Vicuna-7B\cite{zheng2023judging}, Llama-3-8B\cite{meta2024llama3}, Phi-4\cite{phi4}, GPT-3.5\cite{brown2020language}, and GPT-4o\cite{openai2024gpt4o}.

For the retriever-based selection attack, our experimental results are based on a random selection of 20 tools, and the average values are reported. For the LLM-based selection attack, we selected the all tools and averaged the results. We set the number of queries available during the attack process to 10\%. The attack budget is set to $\mathcal{B}_w=\mathcal{B}_c=5000$ for the retriever-based selection and $\mathcal{B}_w=\mathcal{B}_c=2000$ for the LLM-based selection. In the retriever-based attack experiments, \textit{Hit}@{1, 3, 5} are employed as the metrics for tool retrieval, as most current works evaluate the retrieval performance based on the top five retrieved tools \cite{qin2023toolllm, protip, toolrerank}. In the LLM-based attack experiments, we use the tool usage probability as the evaluation metric. We define the usage probability as \( \mathcal{P}_{use} = \frac{1}{|\mathcal{Q}|}\sum_{q\in \mathcal{Q}} \mathcal{E}(q, T) \), where \(\mathcal{E}(\cdot)\) is indicator function, \(\mathcal{E}(q, T)=1\) if target tool was invoked else \(\mathcal{E}(q,T)=0\), and \(\mathcal{Q}\) is the set of queries accessible during the attack process.

\subsection{Retriever-based Attack}
In this setting, we evaluate two types of attacks: \textit{indiscriminate} and \textit{conditional}. 
Indiscriminate attacks force the target tool to compete under any query, while conditional attacks restrict competition to relevant queries with similar tools. 
Results on three retrievers (Table~\ref{tab:G1Res}) show that our method greatly improves the target tool’s recall, enabling weaker retrievers such as BM25 and Ada to achieve competitive performance; for Ada, the Top-3 probability increases by 650$\times$. 
Conditional attacks (Table~\ref{tab:G3Res}) further demonstrate the target tool’s dominance over functionally similar alternatives, highlighting the method’s real-world relevance. 
Overall, our black-box approach reliably elevates the target tool’s ranking and can be optimized for robust cross-query performance.

\subsection{LLM-based Attack}

The overall performance of our attack on LLM-based selection models is summarized in Table~\ref{tab:llmRes}. 
For smaller models such as Llama8b-Instruct and Vicuna-7b, the attack is highly effective, driving the target tool’s invocation probability to nearly 100\%. 
For larger models like GPT-3.5, the probability still increases by 124\%, though perturbations have limited impact on models using in-context learning (e.g., GPT-3.5, GPT-4o). 
We also explored rewriting the target tool description with ChatGPT, but as shown in Table~\ref{tab:rewllm}, the results remained almost unchanged.

\begin{table}[]
\centering
\begin{tabular}{@{}ccc@{}}
\toprule
\multirow{2}{*}[-0.25em]{Model} 
           & \multicolumn{2}{c}{\(\mathcal{P}_{\textit{use}}\)}          \\ \cmidrule(l){2-3} 
           & Origin  & Rewritten  \\ \midrule 
Vicuna-7B
& 0.0519  & 0.0681      \\
Llama-8B
& 0.3491  & 0.3728      \\
GPT-3.5 
& 0.3700  & 0.3219      \\
\bottomrule
\end{tabular}
\caption{The performance of the rewritten tool in LLM-based selection.}
\label{tab:rewllm}
\end{table}

\subsection{Transferability}
We evaluate the transferability of the proposed attack, shown in Tabel \ref{tab:transferretriever} and Table \ref{tab:transferllm}. For retriever-based selection, adversarial texts crafted on BM25 fail to transfer and even degrade performance, likely due to its simple term-matching mechanism. In contrast, adversarial texts generated on Ada transfer effectively, achieving Hit@5 results comparable to direct attacks.
For LLM-based selection, adversarial texts from GPT-3.5-turbo transfer well to Vicuna and Llama, with even stronger transferability observed between Vicuna and Llama. However, texts from these smaller models show limited effectiveness on GPT-3.5. Overall, adversarial texts generated from larger models tend to transfer more easily to smaller ones, while the reverse transfer is less effective.

\subsection{Defense}
To evaluate the stealthiness and robustness of SEEM, we adopt a common defense method in adversarial text attacks—text rewriting—for experimentation, with the results shown in Table \ref{tab:defense}. We observe that while the attack performance experiences some degradation under this defense, it still remains significantly higher than the baseline, thereby ensuring strong attack effectiveness. Furthermore, we employ text perplexity to further analyze the stealthiness of the adversarial texts. The results indicate that the adversarial texts generated by SEEM exhibit an average perplexity increase of less than 50\% compared to the original texts, thus preserving a considerable degree of stealthiness.

\begin{table}[]
\centering
\resizebox*{\linewidth}{!}{
\begin{tabular}{@{}cccccccc@{}}
\toprule
\multirow{2}{*}[-0.25em]{Source Model}  & \multirow{2}{*}[-0.25em]{Target Model}
           & \multicolumn{2}{c}{Hit@1}     & \multicolumn{2}{c}{Hit@3}     & 
           \multicolumn{2}{c}{Hit@5}       \\ \cmidrule(l){3-4} \cmidrule(l){5-6} \cmidrule(l){7-8} 
           & & Origin  & Attack  & Origin  & Attack   & Origin  & Attack    \\ \midrule 

\multirow{2}{*}[-0.25em]{BM25} & Ada
& 12.89  & \cellcolor[HTML]{FFCCC9}6.78  
& 53.39  & \cellcolor[HTML]{FFCCC9}28.36  
& 82.29  & \cellcolor[HTML]{FFCCC9}48.75    \\
&API-retriever
& 12.44  & \cellcolor[HTML]{FFCCC9}5.63  
& 38.64  & \cellcolor[HTML]{FFCCC9}31.65  
& 66.33  & \cellcolor[HTML]{FFCCC9}65.17    \\
\multirow{2}{*}[-0.25em]{Ada} & BM25
& 9.93  & \cellcolor[HTML]{02c39a}47.88  
& 34.11  & \cellcolor[HTML]{02c39a}66.43  
& 57.04  & \cellcolor[HTML]{02c39a}91.76    \\
&API-retriever
& 12.44  & \cellcolor[HTML]{02c39a}33.29  
& 38.64  & \cellcolor[HTML]{02c39a}65.41  
& 66.33  & \cellcolor[HTML]{02c39a}83.33    \\
\multirow{2}{*}[-0.25em]{API-retriever} & BM25
& 9.93  & \cellcolor[HTML]{02c39a}11.55  
& 34.11  & \cellcolor[HTML]{02c39a}52.63  
& 57.04  & \cellcolor[HTML]{02c39a}89.13    \\
&Ada
& 12.89  & \cellcolor[HTML]{FFCCC9}5.42  
& 53.39  & \cellcolor[HTML]{FFCCC9}40.18  
& 82.29  & \cellcolor[HTML]{FFCCC9}58.31    \\
\bottomrule
\end{tabular}
}
\caption{The results of transfer attack across different retriever-based selection model on ToolBench I3. We use red to indicate ineffective transfer attacks and green to indicate effective ones.}
\label{tab:transferretriever}
\end{table}

\begin{table}[]
\centering
\resizebox*{0.7\linewidth}{!}{
\begin{tabular}{@{}cccc@{}}
\toprule
\multirow{2}{*}[-0.25em]{Source Model}  & \multirow{2}{*}[-0.25em]{Target Model}
           & \multicolumn{2}{c}{\(\mathcal{P}_{\textit{use}}\)}     \\ \cmidrule(l){3-4}
           & & Origin  & Attack    \\ \midrule 

\multirow{2}{*}[-0.25em]{Vicuna-7b} & Llama-8B 
& 0.3491  & \cellcolor[HTML]{02c39a}0.6391    \\
&GPT-3.5 
& 0.3700  & \cellcolor[HTML]{FFCCC9}0.3184    \\
\multirow{2}{*}[-0.25em]{Llama-8B} & Vicuna-7b
& 0.0519  & \cellcolor[HTML]{02c39a}0.5731   \\
&GPT-3.5 
& 0.3700  & \cellcolor[HTML]{FFCCC9}0.3349    \\
\multirow{2}{*}[-0.25em]{GPT-3.5} & Vicuna-7b
& 0.0519  & \cellcolor[HTML]{02c39a}0.2193    \\
&Llama-8B
& 0.3491  & \cellcolor[HTML]{02c39a}0.5424    \\
\bottomrule
\end{tabular}
}
\caption{The results of transfer attack across different LLM on ScienceQA. We use red to indicate ineffective transfer attacks and green to indicate effective ones.}
\label{tab:transferllm}
\end{table}

\begin{table}[h]
\centering
\resizebox*{0.6\linewidth}{!}{
\begin{tabular}{@{}cccc@{}}
\toprule
 Model    & Origin  & Attack  &  Defense  \\ \midrule 
API-retriever
& 0.54  & 0.84 & 0.71    \\				
BM25
& 0.34 & 0.89  & 0.65    \\				
Vicuna-7B
& 0.05  & 0.99  & 0.54    \\
Llama-8B
& 0.34  & 0.95  & 0.49    \\ 		
GPT-3.5  
& 0.37  & 0.46  & 0.43    \\
	
\bottomrule
\end{tabular}
}
\caption{Performance changes of SEEM before and after comparing attack and defense, reporting Hit@3 for retriever-based attack and $\mathcal{P}_{use}$ for LLM-based attack.}
\label{tab:defense}
\end{table}

\section{Conclusion}
\label{sec:conclusion}
Our research reveals a significant security issue in
the tool selection stage of Tool Learning. Our experiments demonstrate that slight perturbations to
the tool’s textual information can greatly influence
the tool selection decision. Our findings highlight
the vulnerability of tool selection models and call
for further research and development of more robust models for tool selection.

\section{Acknowledgments}
This work is supported by National Natural Science Foundation of China (62372454, 62576339).

\section{Ethical Statement}
This study investigates manipulation attacks on tool selection to expose security vulnerabilities in tool learning and promote more trustworthy AI systems. While the techniques discussed could be misused, we adhere to responsible disclosure principles, providing only the details necessary for researchers to understand and address these issues without enabling malicious use. All experiments were conducted in controlled settings without real user data, and our work complies with ethical standards to ensure that its contributions serve solely to improve AI security.

\bibliographystyle{IEEEbib}
\bibliography{strings,refs}

\end{document}